# Targeting of chondrocyte plasticity via connexin43 modulation attenuates cellular senescence and fosters a pro-regenerative environment in osteoarthritis

Marta Varela-Eirín[1], Adrián Varela-Vázquez[1], Amanda Guitián-Caamaño[1], Carlos Luis Paíno[2], Virginia Mato[3], Raquel Largo[4], Trond Aasen[5], Arantxa Tabernero[6], Eduardo Fonseca[1], Mustapha Kandouz[7], José Ramón Caeiro[8], Alfonso Blanco[9] and María D. Mayán[1]

## Abstract

Osteoarthritis (OA), a chronic disease characterized by articular cartilage degeneration, is a leading cause of disability and pain worldwide. In OA, chondrocytes in cartilage undergo phenotypic changes and senescence, restricting cartilage regeneration and favouring disease progression. Similar to other wound-healing disorders, chondrocytes from OA patients show a chronic increase in the gap junction channel protein connexin43 (Cx43), which regulates signal transduction through the exchange of elements or recruitment/release of signalling factors. Although immature or stem-like cells are present in cartilage from OA patients, their origin and role in disease progression are unknown. In this study, we found that Cx43 acts as a positive regulator of chondrocyte-mesenchymal transition. Overactive Cx43 largely maintains the immature phenotype by increasing nuclear translocation of Twist-1 and tissue remodelling and proinflammatory agents, such as MMPs and IL-1β, which in turn cause cellular senescence through upregulation of p53, $p16^{INK4a}$ and NF-κB, contributing to the senescence-associated secretory phenotype (SASP). Downregulation of either Cx43 by CRISPR/Cas9 or Cx43-mediated gap junctional intercellular communication (GJIC) by carbenoxolone treatment triggered redifferentiation of osteoarthritic chondrocytes into a more differentiated state, associated with decreased synthesis of MMPs and proinflammatory factors, and reduced senescence. We have identified causal Cx43-sensitive circuit in chondrocytes that regulates dedifferentiation, redifferentiation and senescence. We propose that chondrocytes undergo chondrocyte-mesenchymal transition where increased Cx43-mediated GJIC during OA facilitates Twist-1 nuclear translocation as a novel mechanism involved in OA progression. These findings support the use of Cx43 as an appropriate therapeutic target to halt OA progression and to promote cartilage regeneration.

## Introduction

Osteoarthritis (OA), a chronic disorder characterized by the progressive degradation of articular cartilage, is the most prevalent and disabling musculoskeletal disease worldwide[1,2]. Osteoarthritic cartilage exhibits changes in its extracellular matrix (ECM) composition caused by alterations in component synthesis and degradation[3,4]. Cartilage ECM mainly comprises proteoglycans and collagen type II (Col2A1) and embedded chondrocytes; these

Correspondence: María D. Mayán (Ma.Dolores.Mayan.Santos@sergas.es)
[1]CellCOM research Group, Instituto de Investigación Biomédica de A Coruña (INIBIC), Servizo Galego de Saúde (SERGAS), Universidade da Coruña (UDC), Xubias de Arriba, 84, 15006 A Coruña, Spain
[2]Service of Neurobiology Research, Ramón y Cajal University Hospital (IRYCIS), Madrid, Spain
Full list of author information is available at the end of the article.

cells have low mitotic activity, but high metabolic activity due to their role in ECM remodelling. However, in the early stages of OA, osteoarthritic chondrocytes (OACs) undergo phenotypic changes that increase cell proliferation and cluster formation, with enhanced expression of matrix-remodelling enzymes reflecting attempts to repair the damage. Disruption of the pericellular matrix and progressive cartilage degradation together with changes in subchondral bone, synovial and other joint tissues are characteristic features of disease progression that are associated with increased pain and physical disability[5,6].

The underlying mechanisms of OA are poorly understood and none of the current pharmacological treatments can slow or stop disease progression. However, drugs that promote chondrogenic differentiation in in vitro and in vivo disease models indicate that OACs somehow revert to a less differentiated stage[7–10]. Different molecular hallmarks of OA include the presence of markers of an immature cell phenotype[11–15], suggesting that these cells retain a degree of flexibility[7]. Cell dedifferentiation and reprogramming are associated with wound healing and tissue regeneration[16–18]. Indeed, biological conditions such as tissue injury and ageing promote a precise spatiotemporal cellular plasticity and in vivo reprogramming to achieve tissue repair[19]. However, changes in cell plasticity can also cause pathological processes, such as fibrosis and tumour progression[20–23].

Consistent with other wound-healing diseases, we have found that osteoarthritic cartilage has very high levels of the transmembrane protein connexin43 (Cx43)[24,25]. By coordinating cellular communication through hemichannels (cell-extracellular milieu), gap junctions (GJs; cell–cell) and extracellular vesicles and tunnelling nanotubes, Cx43 plays a key role in many cell functions, including cell proliferation, migration and differentiation in cancer and during development and tissue remodelling[26–29]. Because, Cx43 is involved in wound healing and inflammation, we investigated whether Cx43 might play similar roles during tissue degeneration and repair in OA. Indeed, the normalization of wound healing in skin and heart tissue correlates with Cx43 downregulation at different time points after wounding, which accelerates healing (via modulation of proliferation and migration) and reduces inflammation and fibrosis, promoting a more normal structure with improved mechanical properties[30–33]. Therefore, the chronic overexpression of Cx43 in OA patients due to activation of the wound-healing response may maintain chondrocytes in a more immature (i.e., fibrogenic) phenotype that enables the constant ECM remodelling that leads to cartilage degeneration. This is consistent with (i) Cx43 overexpression from the early stage of the disease[24], (ii) the presence of proliferative chondrocytes in osteoarthritic cartilage and (iii) the typical observation that cartilage is gradually replaced by fibrocartilage with poor biomechanical properties, which leads to joint degeneration[34–39].

The role of connexins in clinical conditions such as inflammation and tissue regeneration has recently gained increased interest. In this study, we focus on how Cx43 affects the phenotype of OACs. Our results reveal a novel role for Cx43 in coordinating cell dedifferentiation and redifferentiation in OA to manage ECM remodelling and to restrain senescence and inflammation. Among the cascade of events leading to OACs reprogramming, upregulation of Cx43 increases the degree of cellular dedifferentiation towards an immature state. Thus, drugs targeting Cx43 to induce redifferentiation may protect against cartilage degradation in OA and boost regeneration.

## Results

### OACs acquire a stem-like phenotype

To test whether immature chondrocytes contribute to OA pathogenesis, we first asked whether chondrocytes isolated from OA patients express stemness-associated cell surface markers (Fig. 1). Using flow cytometry, we characterized the expression of the extracellular markers CD90, CD105, CD29, CD44, CD166 and CD73 during chondrogenic differentiation of human mesenchymal stem cells (hMSCs). As expected, most immature markers decreased during differentiation of hMSCs towards chondrocytes for 7 and 14 days (Fig. 1a). However, high levels of stemness markers such as CD166 were detected on the surface of the OACs (Fig. 1b). In addition, OACs in monolayer culture continued to dedifferentiate to an even more immature phenotype[40,41] correlating with increased levels of the stemness markers CD105, CD44 and CD166 (Fig. 1b and Supplementary Fig. 1). The expression of CD166 indicated that OACs maintained a higher degree of dedifferentiation in long-term culture versus chondrocytes from healthy donors (Fig. 1c). Moreover, CD166 levels were significantly higher in OACs from grade III to IV OA than in those from low-grade OA (Fig. 1d).

To verify this stem-like feature of chondrocytes from OA patients, primary chondrocytes were cultured in chondrogenic medium. We detected a significant reduction in the levels of stemness markers after 14 days in chondrogenic culture (Fig. 1e). In addition, the levels of the transmembrane protein Cx43, previously suggested to be involved in chondrocyte dedifferentiation[42], decreased significantly during growth of OACs in chondrogenic medium for 7 and 14 days (Fig. 1f). The cellular plasticity of OACs was further confirmed using 3D cultures to test cartilage matrix composition and organization. Surprisingly, pellet culture of OACs in chondrogenic medium led to micromasses with characteristic features of healthy cartilage, with high levels of ECM containing proteoglycans and collagen type II (Col2A1) and reduced levels of Cx43 (Fig. 1g and Supplementary Fig. 2a). Pellet culture of healthy chondrocytes in chondrogenic medium improved Col2A1 and proteoglycans synthesis, but to a lesser extent in comparison with OACs (Supplementary Fig. 2b). The reverted phenotype and cellular plasticity acquisition of OACs were confirmed by maintenance of primary cells on chamber slides in osteogenic medium or adipogenic medium (Fig. 1h). Osteogenic and adipogenic differentiation were tested by alizarin red S and oil red O staining, respectively. The results indicated that OACs have high cellular plasticity and ability to differentiate into adipocytes and bone cells (Fig. 1h). However, the ability of healthy chondrocytes to differentiate into adipocytes or bone cells was significantly lower compared with OACs (Fig. 1h). As expected, because these (healthy) chondrocytes were previously expanded in monolayer[43–45], we also detected immature-like characteristics. The findings were consistent with the results shown in Fig. 1b.

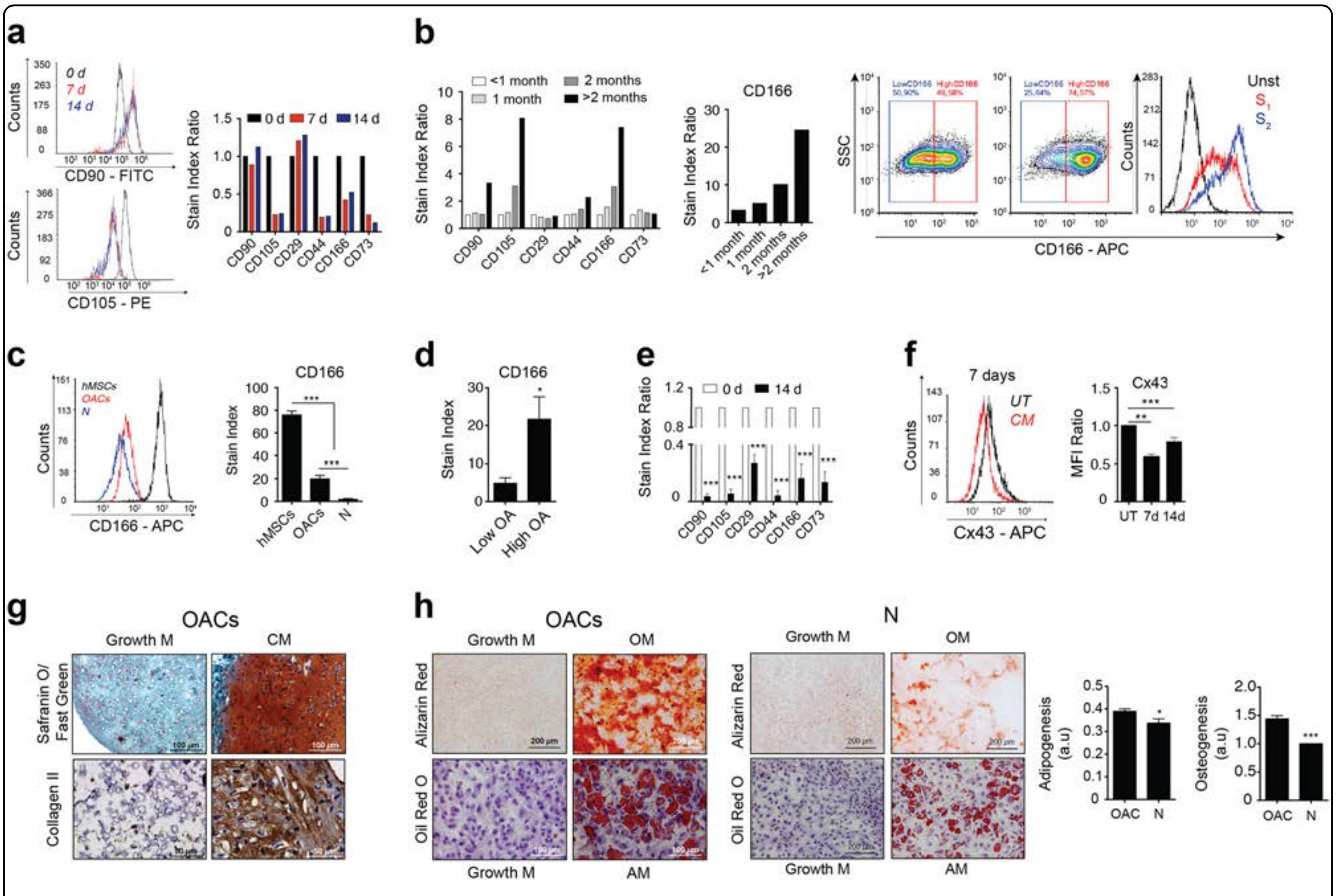

**Fig. 1 OAC reverts to a more progenitor-like or stem cell-like phenotype. a** Mesenchymal stem cell markers (CD90, CD105, CD29, CD44, CD166 and CD73) were tracked by flow cytometry for 7 and 14 days during chondrogenic differentiation of hMSCs in monolayer culture. Data were normalized to the 0-day value. **b** Dedifferentiation of OAC to a more progenitor-like cell state at different time periods in monolayer culture was quantified with the same stem cell markers and flow cytometry. Data were normalized to those of low passage numbers (<1 month; $S_1$). Cell populations of unstained cells (Unst; negative control) and during short (<1 month; $S_1$) and long (>2 months; $S_2$) passages are shown on the right. The heterogeneous subpopulations of dedifferentiated OAC (red line, $S_1$ passage) tended to reach a single population of dedifferentiated chondrocytes cells after long passages (blue line, $S_2$). **c** Comparative flow cytometry analysis showing that OACs express higher levels of the stem cell marker CD166 compared with chondrocytes isolated from healthy donors ($n = 5–8$; mean ± s.e.m.; $^{***}P < 0.0001$; one-way ANOVA). **d** Increased CD166 levels detected by flow cytometry in OACs from high-grade (grades III–IV) versus low-grade (grades I–II) OA donors ($n = 4–5$; mean ± s.e.m.; $^{*}P < 0.05$; Mann–Whitney test). **e** Expression of stem markers by flow cytometry when OACs are cultured in chondrogenic medium (CM) for 14 days suggests that dedifferentiated OACs can redifferentiate into mature chondrocytes ($n = 2$; mean ± s.e.m.; $^{***}P < 0.0001$, Student's *t*-test). **f** Cx43 protein levels in OACs cultured for 7 and 14 days in normal growth medium (DMEM with 10% FBS; UT) or in chondrogenic medium (CM) ($n = 4$; mean ± s.e.m.; $^{*}P < 0.05$; Student's *t*-test). **g** Representative images of OACs that have undergone chondrogenic differentiation (CM) as 3D micromasses for 30 days, and evaluated with Safranin O/Fast Green staining and Col2A1 immunohistochemistry. **h** Comparative images of OACs and healthy chondrocytes (N) that have undergone adipogenic (AM, evaluated by Oil red O staining) or osteogenic (OM, evaluated by alizarin red staining) differentiation for 21 days (AM; adipogenic induction medium, OM; osteogenic medium) OACs show more plasticity and ability to differentiate into adipocytes and bone cells (representative of $n = 10$ experiments; mean ± s.e.m.; $^{*}P < 0.05$, $^{***}P < 0.0001$; Mann–Whitney test). Original magnifications ×20, ×40 and ×100. Scale bars are shown below

### Reduced Cx43-mediated GJIC improves the phenotype of OACs

Cx43 protein is upregulated in OA[24]. We next examined whether GJIC was also altered in chondrocytes from OA patients (Fig. 2a). OACs showed higher levels of Cx43 protein (detected by immunofluorescence, western blot and flow cytometry). Cx43 in primary chondrocytes is localized in appositional membranes as expected for gap junction plaques (Fig. 2a) and OACs showed increased GJIC (detected by LY transfer) compared with healthy chondrocytes (Fig. 2b). Next, we assessed whether intercellular coupling via GJ channels affected the OACs

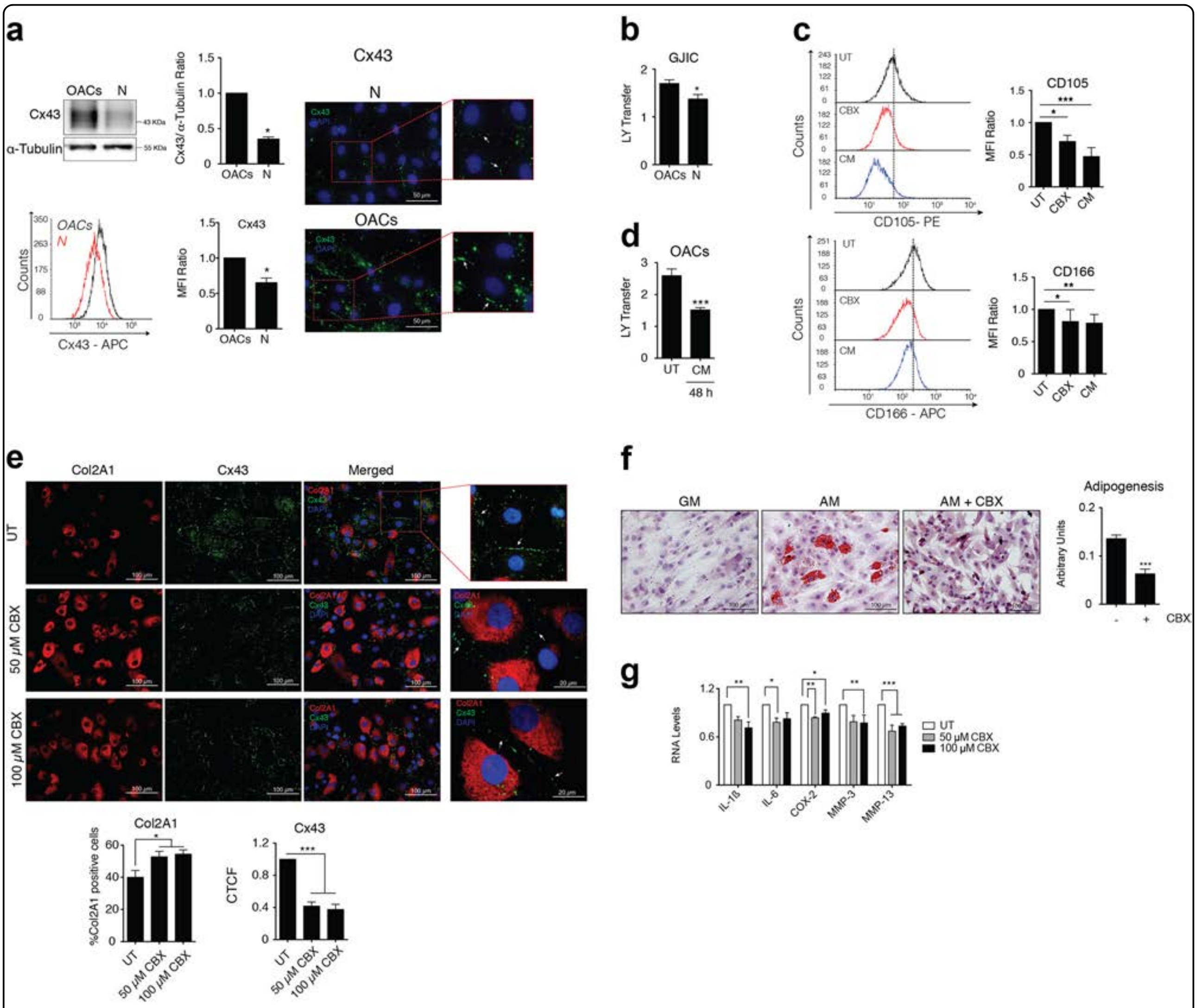

**Fig. 2 Downregulation of Cx43 GJ plaques and GJIC restores chondrocyte redifferentiation in chondrocytes from patients with OA. a** OACs express higher levels of Cx43 protein (mainly located in GJ plaques, white arrows) compared with chondrocytes from healthy donors (N), (western blot, flow cytometry and immunofluorescence assays ($n = 3$; mean ± s.e.m.; $*P < 0.05$; Mann–Whitney test). **b** OACs show higher levels of GJIC in comparison with healthy chondrocytes in primary cultures measured by SL/DT assay ($n = 3$; mean ± s.e.m.; $*P < 0.05$; Mann–Whitney test). **c** Flow cytometry analysis shows reduced levels of CD105 and CD166 in OACs treated for 7 days with 100 μM CBX, suggesting reduced chondrocyte dedifferentiation upon GJIC inhibition. Graphs show CD105 and CD166 levels detected in OACs exposed to CBX or chondrogenic medium (CM) ($n =$ 4–5; mean ± s.e.m.; $*P < 0.05$, $**P < 0.01$, $***P < 0.0001$; one-way ANOVA). **d** OACs in chondrogenic (CM) and normal growth medium (UT, DMEM 10% FBS) for 48 h. GJCI was measured by SL/DT assay ($n = 3$; mean ± s.e.m.; $***P < 0.0001$; Mann–Whitney test). **e** CBX 50 and 100 μM inhibit GJIC in OACs measured by SL/DT assay (Supplementary Fig. 3a). Immunofluorescence analysis of Col2A1 and Cx43 levels in OACs treated with CBX 50 and 100 μM for 24 h. The graphs show the percentage of positive cells for Col2A1 and the CTCF for Cx43 mainly located in the margin of the cells (gap junction plaques, white arrows) ($n = 5$; mean ± s.e.m.; $*P < 0.05$, $**P < 0.01$; one-way ANOVA). **f** Adipogenic differentiation of OACs exposed to CBX 50 μM for 21 days was evaluated with oil red O staining. The graph represents the ratio of cells with lipid deposits to the total number of cells ($n = 8$, mean ± s.e.m.; $***P < 0.0001$; Mann–Whitney test). **g** IL-1ß, IL-6, COX-2, MMP-3 and MMP-13 mRNA expression in OACs treated with 50 and 100 μM CBX for 15 min. Data were normalized to HPRT-1 ($n = 3$–4; mean ± s.e.m.; $*P < 0.05$, $**P < 0.01$, $***P < 0.0001$; one-way ANOVA)

phenotype. Chondrocytes from OA patients were treated with 100 µM of the GJ blocker carbenoxolone (CBX), which inhibits GJIC[46–49] (Supplementary Fig. 3a) or were grown in chondrogenic medium for 7 days (Fig. 2c). Compared with the untreated condition, OACs treated with CBX showed lower levels of the surface stemness markers CD105 and CD106 (Fig. 2c and Supplementary Fig. 3b), and similar to chondrogenic medium conditions, suggesting that the positive effect of Cx43 downregulation on the OACs phenotype also depends on Cx43 channel activity. Interestingly, the chondrogenic medium reduced Cx43 protein levels (Fig. 1f) and GJIC (Fig. 2d).

Next, we sought to confirm whether inhibition of GJIC would modulate chondrocyte plasticity. OACs were grown in monolayer in growth medium supplemented with 50 µM and 100 µM of CBX (Fig. 2e). Consistent with previous results (Fig. 2c), downregulation of GJIC by CBX improved the phenotype of OACs detected by a significant increase in the levels of Col2A1 (Fig. 2e). CBX also decreased Cx43 protein levels at the gap junction plaque as detected by immunofluorescence (Fig. 2e). To further explore the effect of Cx43 plaques and GJIC on cell plasticity, OACs were grown in adipogenic medium in the presence of 50 µM CBX. Notably, inhibition of GJIC by CBX significantly decreased the adipogenic differentiation of OACs as detected by oil red O staining (Fig. 2f). However, CBX treatment increased the osteogenic differentiation of OACs when these cells were grown in osteogenic medium with 50 µM CBX for 14 days (Supplementary Fig. 4). Consistent with this finding, previous studies have suggested a direct transformation of mature chondrocytes into osteoblasts and osteocytes[50–52], reemphasizing the role of Cx43 as a modulator of cell plasticity in chondrocytes, whose downregulation enhances OACs redifferentiation to mature chondrocytes. As expected, redifferentiation of OACs by reducing Cx43 plaques and GJIC with CBX (Fig. 2c, e) led to reduced expression of the proinflammatory mediators IL-1ß, IL-6 and COX-2 and reduced levels of MMP-3 and MMP-13 (Fig. 2g).

### Cx43 promotes phenotypic reversion by modulating Twist-1

Transfection of the T/C-28a2 human chondrocyte cell line with a Cx43 overexpression vector significantly increased Cx43 protein levels, Cx43 membrane localization, GJIC (Fig. 3a) and levels of the stemness surface markers CD166 and CD105 (Fig. 3b). Similarly, reduction of Cx43 levels in the T/C-28a2 human chondrocyte cell line by CRISPR/Cas9-mediated heterozygous Cx43 gene knockout led to a downregulation of the GJIC (Fig. 3c) and the stemness surface markers CD166 and CD105 (Fig. 3d).

The reversion of the cell phenotype in OACs can be regulated by several mechanisms, including transcription factors involved in the epithelial-to-mesenchymal transition (EMT) such as Twist-1. In addition, Twist-1 is a regulator of MSCs differentiation into chondrocytes[53] and is expressed in immature chondrocytes but decreased in mature chondrocytes[54]. In addition, Twist-1 is upregulated in human OA knee cartilage[55]. Significantly, overexpression of Cx43 in chondrocytes led to nuclear translocation of Twist-1, as detected by western blot, together with an increase in nuclear PCNA, which is associated with cell proliferation (Fig. 3e). The increase in Twist-1 was accompanied by an elevated expression of the EMT markers N-cadherin and vimentin (Fig. 3e) and with changes in the vimentin cytoskeleton and cell shape (Fig. 3f). Notably, the mRNA expression levels of N-cadherin, vimentin and Twist-1 were significantly reduced when Cx43 was downregulated in the heterozygous Cx43 gene knockout cell line. Downregulation of Cx43 prevented from nuclear localization of Twist-1 (Fig. 3g).

### Downregulation of Cx43 protects OACs from senescence and inflammation

Dedifferentiation can be induced by various stress conditions and has the potential to accelerate ageing processes[56,57]. Moreover, increased cellular senescence was recently reported in OA cartilage[58]. In accordance with these results, we observed that high levels of Cx43 were correlated with increased cellular senescence in human chondrocytes (T/C-28a2), as detected by the senescence-associated beta-galactosidase (SAβG) activity measured by flow cytometry (Fig. 4a). To verify that altered Cx43 expression leads to senescence, we studied changes in the replicative senescence factor p53 and its signalling partner $p16^{INK4a}$. Gene expression and western blot analysis revealed increased levels of p53 and $p16^{INK4a}$ when Cx43 was overexpressed (Fig. 4b). In addition, cellular senescence and the senescence-associated secretory phenotype (SASP) can convert senescent cells into proinflammatory cells. Increased levels of Cx43 were sufficient to induce higher expression of proinflammatory mediators involved in cartilage degradation in OA, such as IL-1β, COX-2 and the metalloproteinases 3 and 13 (Fig. 4c). As expected, upregulation of Cx43 activated NF-κß (nuclear translocation), a master regulator of the SASP (Fig. 4d). Hence, a reduction in Cx43 levels via the CRISPR/Cas9 approach (T/C-28a2 cells heterozygous for Cx43) decreased the levels of nuclear NF-κß (Fig. 4d), together with a significant reduction in the levels of chondrocyte senescence detected by SAβG activity and $p53/p16^{INK4a}$ levels (Fig. 4e). Downregulation of Cx43 significantly reduced the synthesis of the inflammatory and cell reprogramming factors IL-1ß and IL-6 and the ECM remodelling enzymes MMP-3 and MMP-13 (Fig. 4f). Taken together, these results suggest that overexpression of Cx43 contributes to dedifferentiation of chondrocytes in OA (chondrocyte-mesenchymal transition via Twist-1), but also favours an inflammatory environment and premature senescence, which in the long term may contribute to excessive tissue remodelling and degradation.

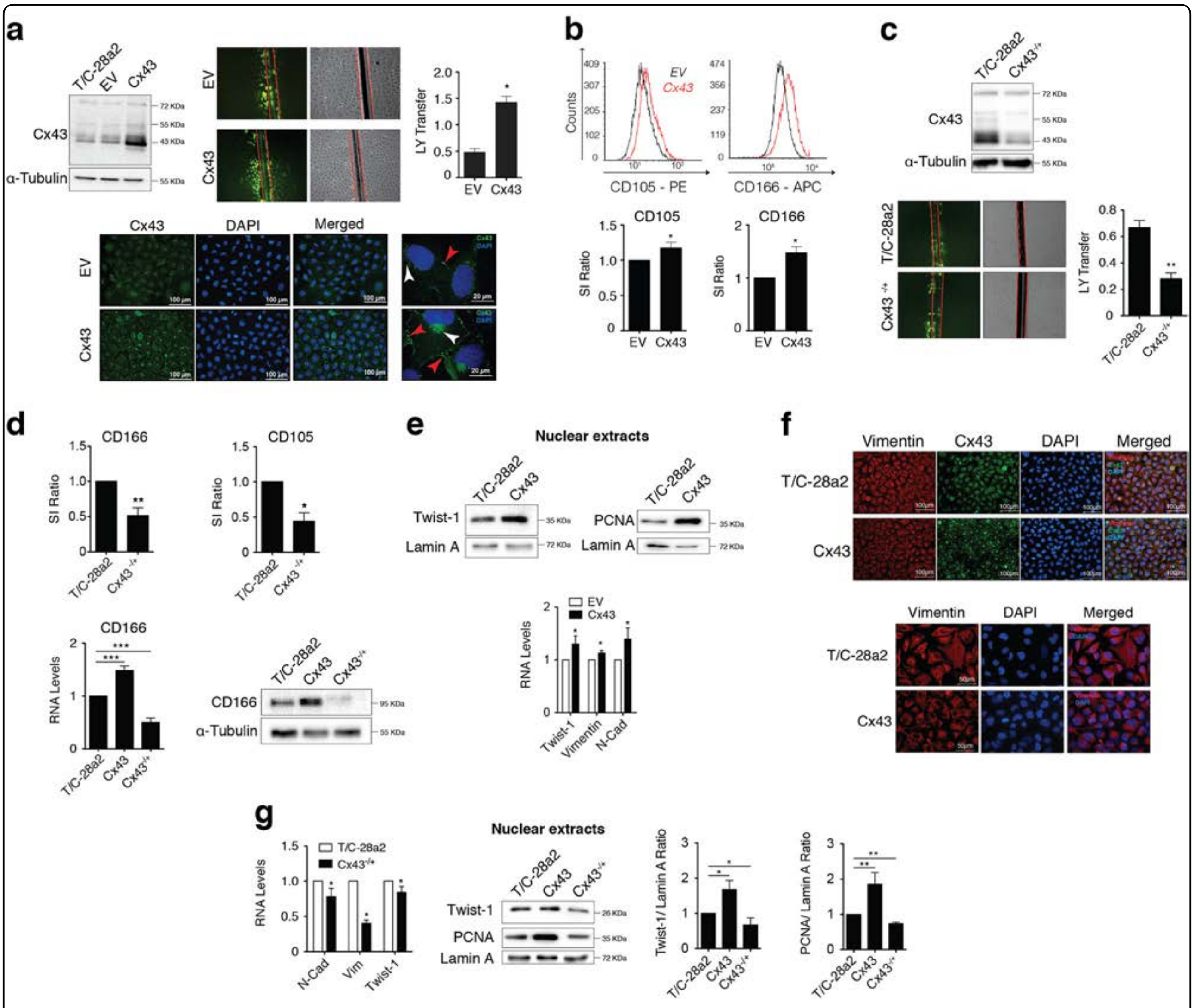

**Fig. 3 Upregulation of Cx43 induces Twist-1 activation and switches on the dedifferentiation program to gain mesenchymal characteristics.** **a** The T/C-28a2 chondrocyte cell line transfected with a plasmid to overexpress Cx43 (referred to as Cx43) displays increased Cx43 protein levels (detected by western blot) and GJIC (evaluated by LY transfer in an SL/DT assay). Cx43 levels were compared with those of the non-transfected and empty vector (EV)-transfected cells. The graph represents the quantification of more than 10 images from two independent experiments (*$P < 0.05$; Mann–Whitney test). Below, immunofluorescence assays indicating that, in chondrocytes overexpressing Cx43, Cx43 (green) is mainly localized to the membrane (red arrows) and perinuclear area (white arrows). Nuclei were stained with DAPI. **b** Flow cytometry analysis showing that T/C-28a2 cells overexpressing Cx43 (red) expressed significantly higher levels of the stemness markers CD105 ($n = 7$) and CD166 ($n = 5$) versus the same cell line transfected with a control plasmid (black) (mean ± s.e.m.; *$P < 0.05$; Mann–Whitney test). **c** Downregulation of Cx43 levels in T/C-28a2 cells with the CRISPR-Cas9 knockdown system (Cx43$^{-/+}$) (detected by western blot) led to significantly lower levels of GJIC, detected by SL/DT assay ($n = 6$, mean ± s.e.m.; **$P < 0.01$; Mann–Whitney test). **d** Compared with the control cells, the CRISPR-Cas9 knockdown chondrocyte cell line (T/C-28a2, Cx43$^{-/+}$) contained lower levels of the stem cell markers CD105 (flow cytometry) and CD106 (flow cytometry, western blot and RT-qPCR). Flow cytometry; $n = 7$ mean ± s.e.m.; *$P < 0.05$, **$P < 0.01$; Mann–Whitney test. Gene expression was normalized to HPRT-1 levels ($n = 7$, mean ± s.e.m.; ***$P < 0.0001$; one-way ANOVA). **e** Upregulation of Cx43 increases the levels of nuclear Twist-1 and nuclear PCNA (western blot) and upregulates the gene expression of Twist-1 and the mesenchymal markers vimentin and N-cadherin (RT-qPCR, below) ($n = 3$–$5$; mean ± s.e.m, *$P < 0.05$; Mann–Whitney test). T/C-28a2 cell line overexpressing Cx43 compared with non-transfected cells or control plasmid (EV). Lamin A was used as a nuclear loading control. **f** Changes in the vimentin (red) filament network were associated with Cx43 upregulation. T/C-28a2 cells overexpressing Cx43 versus non-transfected cells. Nuclei were stained with DAPI. Scale bar is shown below. **g** Reduced levels of Cx43 in T/C-28a2 cells (CRISPR-cas9 system) correlated with downregulation of N-cadherin, vimentin and Twist-1 gene expression ($n = 3$, mean ± s.e.m.; *$P < 0.05$; Mann–Whitney test). Cx43 levels correlated with nuclear levels of Twist-1 and PCNA, detected by western blot. T/C-28a2 cell line compared to the same cells overexpressing Cx43 (Cx43) or the heterozygous for Cx43 (Cx43$^{-/+}$). Lamin A was used as a nuclear loading control ($n = 2$–$3$, mean ± s.e.m.; *$P < 0.05$, **$P < 0.01$; Student's *t*-test)

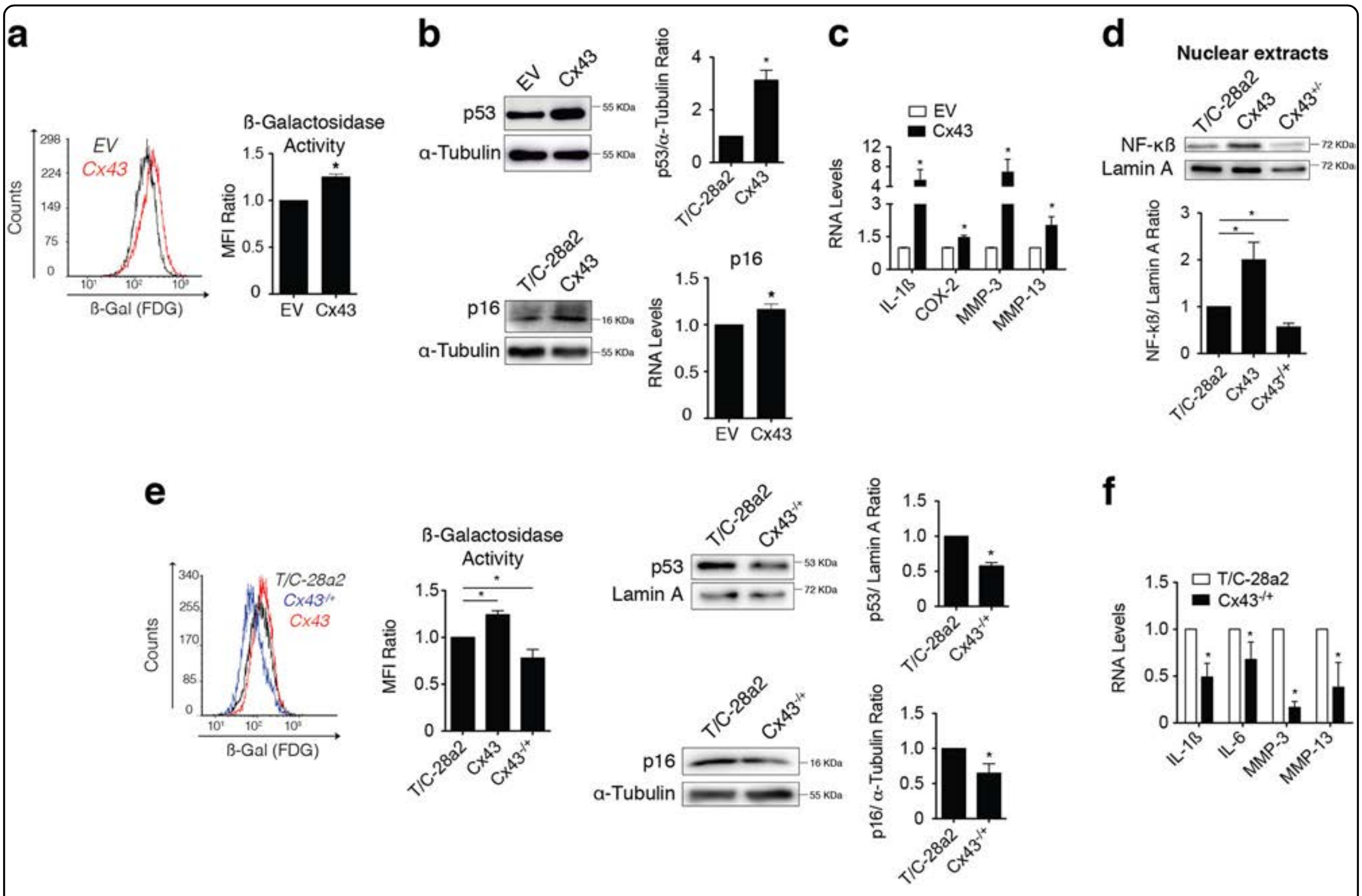

**Fig. 4 Dedifferentiation induced by Cx43 activates senescence and prolongs inflammation and remodelling. a** Flow cytometry analysis with the FDG substrate showing elevated levels of SAβG activity in Cx43-overexpressing T/C-28a2 cells (red) versus empty vector (EV) control cells (black) ($n = 5$; mean ± s.e.m.; *$P < 0.05$; Mann–Whitney test). **b** Western blot ($n = 2$; mean ± s.e.m.; *$P < 0.05$; Student's *t*-test) and RT-qPCR ($n = 4$; mean ± s.e.m.; *$P < 0.05$; Mann–Whitney test) showing that overexpression of Cx43 increased p53 and p16 levels in T/C-28a2 chondrocytes. **c** Analysis of IL-1β, COX-2, MMP-3 and MMP-13 mRNA levels using real-time PCR in T/C-28a2 cells overexpressing Cx43 ($n = 2–4$; normalized to HPRT-1 and represented as mean ± s.e.m.; *$P < 0.05$; Mann–Whitney test). **d** Nuclear levels of NF-κB detected by western blot in the T/C-28a2 cell line overexpressing Cx43 (Cx43) and in the T/C-28a2 heterozygous for Cx43 ($Cx43^{-/+}$), compared to the non-transfected cell line (T/C-28a2). Lamin A was used as a nuclear loading control ($n = 2$, mean ± s.e.m.; *$P < 0.05$, **$P < 0.01$; Student's *t*-test). **e** Downregulation of Cx43 in T/C-28a2 cells (CRISPR-Cas9 knockdown) decreased SAβG (blue) ($n = 3–4$; mean ± s.e.m. *$P < 0.05$, **$P < 0.01$; one-way ANOVA), and increased of p53 and p16 protein levels, detected by western blot ($n = 2–3$, mean ± s.e.m.; *$P < 0.05$; Student's *t*-test). **f** Analysis of IL-1β, COX-2, MMP-3 and MMP-13 mRNA levels in the T/C-28a2 heterozygous for Cx43 ($Cx43^{-/+}$). Data were normalized to HPRT-1 (mean ± s.e.m. $n = 3–4$; *$P < 0.05$; Mann–Whitney test)

## Discussion

During OA, chondrocytes undergo dedifferentiation and senescence, but how these processes are activated and coordinated is poorly understood. Herein, we demonstrate that the GJ channel protein Cx43 acts as a positive regulator that reverts chondrocytes to a less differentiated state, possibly by upregulating the activity of factors such as the basic helix-loop-helix transcription factor Twist-1 (Fig. 3) and inflammatory cytokines such as IL-1β (Fig. 4). Overexpression of Cx43 and enhancement of GJIC in OACs compromises the ability of these cells to redifferentiate, promoting (and prolonging) a stem-like state by activating chondrocyte-mesenchymal transition. However, this is a reversible loss, because downregulation of Cx43 and GJIC allow chondrocytes to properly execute the redifferentiation program. In previous studies, OACs were shown to have a compromised ability to redifferentiate in monolayer and 3D culture[59,60]. Our findings suggest a new paradigm in which the dedifferentiation and redifferentiation of chondrocytes are regulated by a Cx43-sensitive circuit. Similarly, in an intracerebral haemorrhage mouse brain model Cx43 was recently shown to regulate a phenotypic switch by activation astroglial-mesenchymal transition via nuclear translocation of the Yes-associated protein (YAP), a potent transcription coactivator[61]. Cx43 was found to sequester YAP at gap junction plaques, and downregulation of Cx43 caused Cx43-YAP complex dissociation and nuclear translocation of YAP[61].

Although progenitor-like cells have been found in different zones of adult cartilage, the origin and functions of these cells are poorly understood. Based on multilineage potential and stemness markers, it has been accepted that these cells act as stem cells within the cartilage[62–64]. Further findings indicated the emergence of stem-like cells in healthy cartilage explants in response to blunt mechanical impact or scratch injury[65]. In addition, these stem-like cells show a distinct phenotype versus hMSCs isolated from the same donors[65]. Additional evidence strongly suggested that the phenotype of these stem cells is correlated with the degree of ECM degradation in cartilage from OA patients[66,67]. Our data support the hypothesis that these cells, more than stem/progenitor cells (i.e., hMSCs), are chondrocytes dedifferentiated towards a stem-like state via EMT. In accordance with our results, CD166-positive chondrocytes have strong chondrogenic potential[15]. In addition, chondrocytes isolated from cartilage have chondrogenic, osteogenic and adipogenic differentiation potential (Fig. 1)[11,68].

Here, we show that overexpression of Cx43 is sufficient to increase the nuclear translocation of the EMT transcription factor Twist-1 in the established healthy human chondrocyte cell line T/C-28a2, together with an increase in vimentin, N-cadherin and the stemness markers CD105 and CD166 (Fig. 3). The phenotypic change is accompanied by increased expression of metalloproteinases (MMP-13 and MMP-3) and proinflammatory mediators (IL-1β and COX-2) (Fig. 4). Hence, downregulation of Cx43 or GJIC in chondrocytes (via targeted heterozygous deletion using CRISPR technology and CBX treatment) reduces stemness markers and improves Col2A1 synthesis (Fig. 2) together with a decrease in the expression of metalloproteinases and proinflammatory mediators (Fig. 4). Our demonstration that downregulation of Cx43 can rescue the differentiation of OACs may provide important insights into the conflicting differentiation status during OA and aid in the development of effective therapeutic strategies to stop or reverse joint damage.

Although the mechanisms by which Cx43 regulates OACs dedifferentiation/redifferentiation require further investigation, we observed that Cx43 overexpression increased the nuclear levels of Twist-1, leading to increased cell proliferation and upregulation of the EMT markers vimentin and N-cadherin (Fig. 3). Following injury, the dedifferentiation and redifferentiation processes play important roles in the generation of proliferating regenerative cells with the ability to restore the lost tissue in a precise way[69,70]. Interestingly, Twist-1 has been linked to muscle regeneration through dedifferentiation by downregulating myogenin gene expression and reversing cell differentiation in the absence of growth factors[71]. Undoubtedly, understanding of how Cx43 coordinates tissue regeneration would have therapeutic implications for cartilage conditions, as has been shown for skin wound healing[72–74] and bone remodelling[75]. Co-treatment of OACs with the GJIC-inhibitor CBX affected the redifferentiation capacity of OACs (Fig. 2), indicating that connexin channels and GJIC participate in the activation and nuclear translocation of Twist-1 and the expression of cytokines. However, inhibition of GJIC by CBX also decreased Cx43 plaques (Fig. 2e and Supplementary Fig. 5). Together, our results on the inhibition of dye transfer and downregulation of Cx43 in CBX-treated chondrocytes suggest that chondrocyte redifferentiation observed upon CBX treatment occurred through a GJ-dependent manner (Cx43 plaques). However it is difficult to exclude Cx43 GJ-independent mechanisms in these phenotypic switching because CBX also reduced the levels of Cx43. In fact, and in concordance with our results, it has been recently reported that a smaller carboxy-terminal fragment of Cx43 (Cx43-20K) is a transcriptional regulator of N-cadherin by interacting with the RNA polymerase II and with several transcription factors, including BTF3 at the N-cadherin gene promoter region[76]. Importantly, it has also been reported that Cx43 modulates gene transcription by altering the recruitment of the Sp1/Sp3 transcription factors to CT-rich connexin-response elements (CxREs) within promoters[77,78]. The promoters of Twist-1 and Col2A1 have Sp1-binding sites and are modulated by the Sp1/Sp3 complex[79–81]. Notably, reduction in Cx43 at the membrane and attenuation of GJIC in a rat osteoblast-like osteosarcoma cell line (ROS 17/2.8) resulted in reduced Sp1 recruitment to CxREs through decreased Sp1 phosphorylation by the ERK mitogen-activated protein kinase (MAPK) cascade[78]. Additionally, novel loci recently associated with OA in a large GWAS include components such as MAP2K6 on chromosome 17, which is related to the MAPK pathway[82], which can interfere with Cx43 trafficking, stability, channel function and signalling[83,84].

As mentioned above, chondrocytes in culture undergo dedifferentiation via EMT and senescence. Increased senescence was recently observed in mice after traumatic OA, peaking 2 weeks after injury[58]. While dedifferentiation is seen in early passages, a higher number of senescent cells are observed in late passages compared with early passages[85]. Transplantation of such dedifferentiated and senescent cells results in the formation of unwanted fibrous cartilage at the transplantation site[86,87]. The identification of the mechanisms involved in in vitro and in vivo redifferentiation to a chondrocyte-specific phenotype such as modulation of Cx43 function is needed to efficiently develop these therapies. On the other hand, senescent cells have also been suggested to be a promising target for improving cartilage regeneration after articular joint injury in mice[58]. Indeed, treatment of human OACs in monolayer and 3D pellet culture with the senolytic drug UBX0101 for 4 days improves chondrocyte-specific properties, increasing the levels of COL2A1 and proteoglycans and decreasing the OA-related genes MMP-13 and IL-1β and the senescence-related genes MMP-3 and IL-6[58]. Here, we show that downregulation of Cx43 also decreases the propensity of chondrocytes to undergo cellular senescence. Accordingly, upregulation of Cx43 leads to senescence, along with upregulation of p53/p16, which has been associated with chondrocyte senescence[58,88], and to nuclear translocation of NF-κB, the master regulator of SASP. In this context, in the same way downregulation of Cx43 prevents senescence, NF-κB activation and cytokine/MMPs gene expression (Fig. 4). IL-1α and IL-1β, through cell surface receptors, act as upstream regulators of NF-κB and SASP by activating numerous genes, including those encoding IL-6, IL-8 and proinflammatory cytokines that reinforce senescence[89–94]. Notably, Cx43 overexpression alone in chondrocytes leads to a fivefold increase in IL-1β gene expression, which is a catabolic factor in OA (Fig. 4c)[95–97]. Indeed, IL-1β upregulates Cx43 in chondrocytes and synoviocytes from patients with OA and rheumatoid arthritis[98–100], as well as p53[101,102] and NF-κB[90,103]. Thus, IL-1β may act as a feedback loop factor to amplify the effect of Cx43 on senescence and chondrocyte phenotypic changes in OA. Among these feedback loop amplification events, it is important to note that the Twist-1 promoter has a functional NF-κB binding site and that overexpression of NF-κB is sufficient to induce transcriptional upregulation of Twist-1, along with EMT and stemness properties[104].

Transient induction of cell senescence as part of a damage response orchestrates tissue remodelling and repair by promoting dedifferentiation, proliferation and ECM remodelling[105–107]. Dysregulated wound healing caused by excessive signalling and accumulation of senescent cells, resulting from either excessive production or defective clearance, acts as a barrier to effective repair[107–109]. SASP factors convert senescent cells into proinflammatory cells[110]. Importantly, IL-6 and other soluble factors that strongly promote cellular plasticity by activating cellular reprogramming in nearby non-senescent cells[105,107,109,111] probably halt redifferentiation by potentiating the chronic dedifferentiation of chondrocytes in OA. In fact, it has recently been reported that targeting of the gp130 receptor for the IL-6 family and activation of ERK and NF-κB improves cartilage healing in an animal model of OA[112]. Our discovery that Cx43 regulates dedifferentiation/redifferentiation, in addition to senescence, highlights an important intersection between chondrocyte reprogramming and wound healing relevant for understanding the molecular basis of cartilage degeneration in OA, which offers opportunities for new therapies.

OACs in the cartilage of patients with early- and late-stage OA are characterized by Cx43 overexpression. Our study shows that Cx43 causes chondrocyte-mesenchymal transition and senescence that culminate in cartilage degeneration. Mechanisms that establish redifferentiation (chondrocyte-specific phenotype), such as Cx43 downregulation, will potentially restore cartilage repair in OA and improve our ability to implement cell therapy approaches for cartilage diseases. The present work demonstrates that control of cellular plasticity and senescence in chondrocytes via the single master factor Cx43, which upregulates Twist-1/IL-1β and p53/p16-NF-κB-IL-6 signalling pathways (Fig. 5), may be an attractive candidate for tissue engineering and regenerative medicine strategies for OA treatment and for promotion of proper cartilage repair.

## Methods

### Tissue collection, processing and cell culture

Cartilage collection and processing were performed as previously described[25]. The study was conducted with the approval of the institutional ethics committee (C.0003333, 2012/094 and 2015/029) and after informed consent were obtained. We selected samples from healthy and from moderate OA (grades I–II) and high OA (grades III–IV) groups. OA samples were assigned based on individuals' medical record data and histological analysis as previously described[24]. Healthy individuals suffered a knee or hip fracture and had no history of joint disease. Normal tissue was studied by histological analysis[24]. Cartilage was obtained in situ from 35 donors including healthy and OA patients (grades I–II and grades III–IV). For the isolation and culture of primary chondrocytes, fresh cartilage was rinsed with saline and cells were isolated as previously described[113]. Cells ($2 \times 10^6$) were then seeded into 100-mm dish plates and incubated at 37 °C with 5% $CO_2$ and 100% humidity in Dulbecco's Modified Eagle's medium (DMEM) supplemented with 100 U/ml penicillin, 100 μg/ml streptomycin and 10% foetal bovine serum (FBS; all from Gibco, Thermo Fisher Scientific) until ~80–90% confluence was reached. The T/C-28a2 chondrocyte cell line, kindly donated by Dr. Goldring (the Hospital for Special Surgery, New York, USA), was cultured in DMEM supplemented with 10% FBS, 100 U/ml penicillin and 100 μg/ml streptomycin. hMSCs were obtained with informed consent from bone marrow donors recruited by the Bone Marrow Transplant Programme of the Haematology Service at the Hospital Universitario Reina Sofía (Córdoba, Spain) and from subcutaneous inguinal fat obtained with signed informed consent from healthy individuals who underwent catheter insertion at the Hospital Universitario Ramón y Cajal (Madrid, Spain).

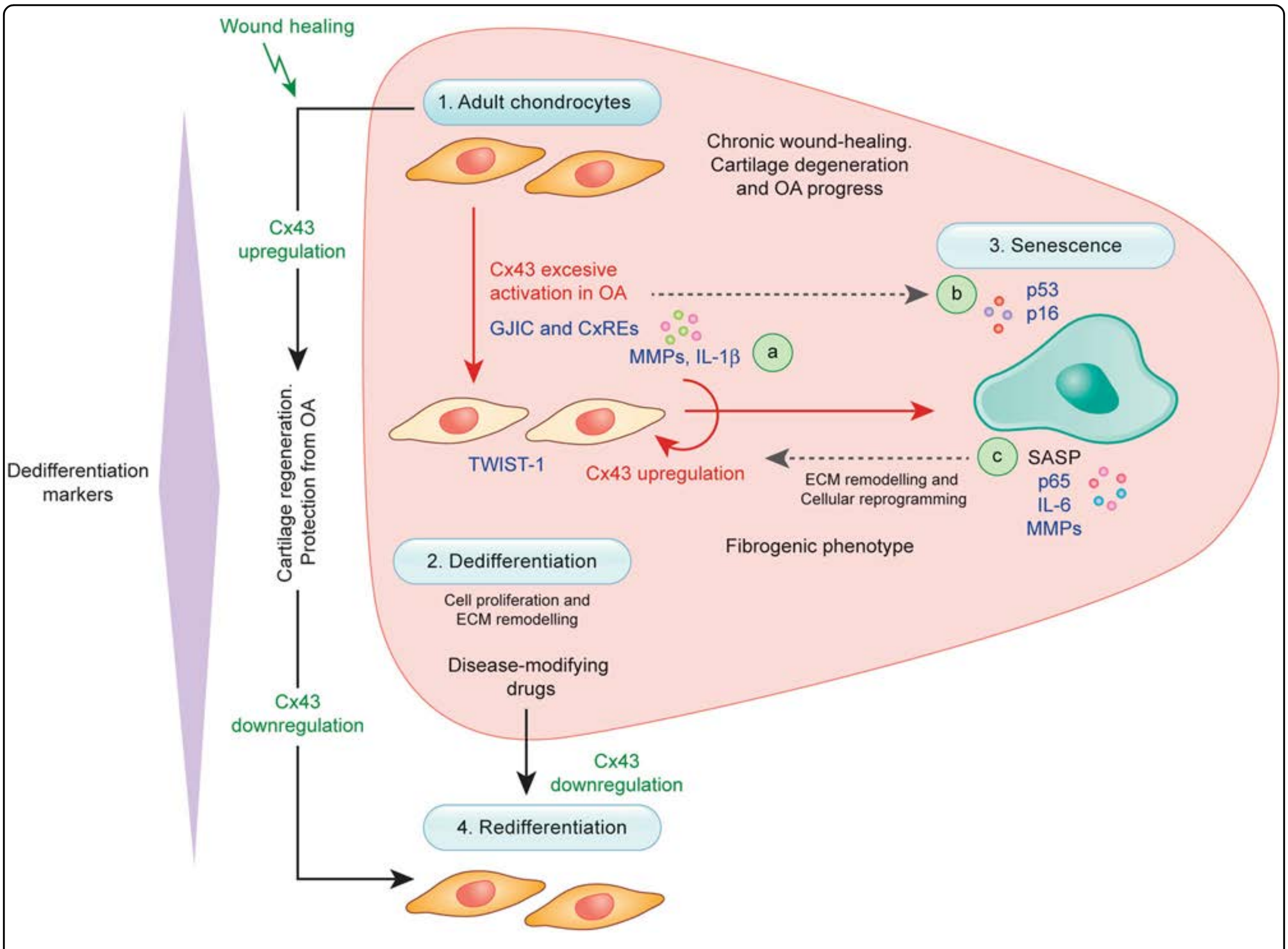

**Fig. 5 Downregulation of Cx43 in OA fosters a pro-regenerative environment by favouring redifferentiation.** Accumulation of dedifferentiated (2) and senescent (3) chondrocytes and Cx43 overexpression are typical in the cartilage microenvironment from the early stages of OA. Phenotypic changes associated with upregulation of Cx43 during tissue repair may contribute to ECM remodelling and cartilage regeneration (or wound healing) (1–2 and 4, black arrows). However, they result in a chondrocyte fibrogenic phenotype (2, and red arrow) and accumulation of senescent cells (3) in OA, which is unable to support the ECM composition by continuous activation of ECM remodelling (MMPs) and proinflammatory factors, such as IL-1ß, NF-κB and IL-6 **a**, **c**. Downregulation of Cx43 or GJIC (e.g., by CRISPR-Cas9 or CBX treatment) leads to redifferentiation of OACs (4) by reducing Twist-1 activation (2), senescence (3 and **b**) and SASP (**c**, NF-κB-p65). The results indicate that dedifferentiation and senescence in OA are flexible processes that can be reverted by modulating Cx43. The elimination or inhibition of factors that contribute to the prolongation of cellular dedifferentiation (Twist-1; chondrocyte-mesenchymal transition) or reprogramming (e.g. via IL-6) reduces the burden of dedifferentiated (2) and senescent (3) cells, which in turn decreases the synthesis of catabolic factors (**a** and **c**) involved in cartilage degradation in OA and leads to a mature phenotype able to support the cartilage ECM composition

hMSCs were cultured in α-minimum essential medium (α-MEM; Lonza) supplemented with 10% FBS, 100 U/ml penicillin, 100 µg/ml streptomycin, 2 mM GlutaMax (Gibco, Thermo Fisher Scientific) and 1 ng/ml recombinant human fibroblast growth factor-2 (rhFGF-2; Immunotools) or in MesenPRO RS$^{TM}$ Medium supplemented with 100 U/ml penicillin and 100 µg/ml streptomycin.

## Adipogenic differentiation

Confluent cells were incubated with a commercially available adipogenic medium (AM, hMSC Adipogenic Differentiation Bullekit$^{TM}$, Lonza) for 21 days, changing the medium every 2–3 days. Adipogenic differentiation was evaluated by the oil red O staining. Briefly, lipid droplets in cells were stained for 20 min in freshly prepared and filtered 60% (v/v) oil red O solution prepared from a 0.5 mg/ml stock (Sigma-Aldrich). Then, cells were rinsed with 60% (v/v) isopropanol, followed by a final rinse in $dH_2O$. Slides were imaged using an Olympus BX61 microscope and a DP71 digital camera (Olympus). Lipid droplet-containing cells were counted with ImageJ software version 1.48 and adipogenic differentiation was evaluated as the ratio of positive droplet-containing cells to the total number of cells.

## Osteogenic differentiation

Chondrocytes or hMSCs were cultured until confluent and differentiated for 21 days with a commercial osteogenic medium (StemPro® Osteogenesis Differentiation Kit, Gibco, Thermo Fisher Scientific). The medium was changed every 2–3 days. Osteogenesis was evaluated by calcium deposit formation with alizarin red S staining. Acetone-fixed cells were stained with a 2% (w/v) alizarin red S solution (Sigma-Aldrich) for 5 min at room temperature (RT) and rinsed with $dH_2O$. Slides were imaged using an Olympus BX61 microscope and a DP71 digital camera (Olympus). Alizarin red positivity was analysed using ImageJ software version 1.48.

## Chondrogenic differentiation

Chondrogenesis was induced in chondrocytes or hMSCs with a commercial chondrogenic medium (StemPro® Chondrogenesis Differentiation Kit, Gibco, Thermo Fisher Scientific). For flow cytometry analysis of chondrogenesis, cells were cultured as a monolayer for 7 and 14 days in the presence of chondrogenic medium. A total of 25,000 cells seeded into 8-well culture chambers were used for chondrogenesis, osteogenesis and adipogenesis assays. The differentiation in monolayer started when cells reached 90% confluence (after 2 days in culture). For immunohistochemical analysis, cells were cultured as micromasses (500,000 cells per pellet) for 30 days in the presence of chondrogenic medium. Micromasses were embedded in Tissue-Tek® OCT™ (Sakura) and chondrogenesis was evaluated with the ECM-specific stains toluidine blue, safranin O-fast green and Col2A1 immunohistochemistry.

## Western blot

Cells were lysed in ice-cold lysis buffer (150 mM NaCl, 50 mM Tris-HCl, pH 7.5, 5 mM EDTA, pH 8, 0.5% v/v Nonidet P-40, 0.1% (w/v) SDS, 0.5% (v/v) sarkosyl) supplemented with 5 μg/ml protease inhibitor cocktail and 1 mM phenylmethylsulfonyl fluoride (PMSF; Sigma-Aldrich). For nuclear protein isolation, the NE-PER™ kit (Thermo Fisher Scientific) was used according to the manufacturer's instructions. Total protein content was determined by a Bradford protein assay. Ten micrograms of total protein was separated on 10% SDS-PAGE and transferred to a polyvinylidene fluoride (PVDF) membrane (Millipore Co., Bedford, MA). Protein transfer was checked by staining the membrane with ATX Ponceau S red staining solution (Sigma-Aldrich). The membrane was then blocked with 5% milk in Tris-buffered saline (TBS; 20 mM Tris, 150 mM NaCl) and 0.05% Tween-20 (Sigma-Aldrich) and incubated with primary antibody overnight at 4 °C; HRP-secondary antibody incubation was performed at RT for 1 h. The signal was developed using Pierce™ ECL Western Blotting Substrate in an LAS-300 Imager (Fujifilm). The following primary antibodies were used: α-tubulin (Sigma-Aldrich, T9026), Cx43 (Sigma-Aldrich, C6219), Twist-1 (Santa Cruz Biotechnology, sc-81417), PCNA (Santa Cruz Biotechnology, sc-56), CD166 (Santa Cruz Biotechnology, sc-74558), p16$^{INK4a}$ (Abcam, ab108349), p53 (Santa Cruz Biotechnology, sc-126), NF-κB (Santa Cruz Biotechnology, sc-8008) and lamin A (Santa Cruz Biotechnology, sc-20680).

## Scrape loading/dye transfer (SL/DT) assay

An SL/DT assay was performed to evaluate GJIC as previously described with modifications[25]. Cells were seeded on 12-well culture plates and cultured until 70%–100% confluent. Cells were rinsed twice with warm PBS and a 0.4% (w/v) solution of lucifer yellow (LY) (Cell Projects Ltd©, Kent, UK) in PBS was loaded. Then, two distant scrapes were made across the culture plate with a 29-gauge needle and scalpel and the cells were allowed to take up the dye for 3 min at 37 °C. After being washed twice with FBS-free DMEM, the cells were fixed with 4% formaldehyde and dye transfer was evaluated in a Nikon Eclipse Ti fluorescent microscope with a ×10 objective lens. The number of dye-positive cells (LY transfer) from the cut site to the farthest detectable uptake of LY reflects the GJ connectivity between cells. The score was calculated as previously reported[114,115].

## Antigen expression analysis by flow cytometry

For the detection of membrane-associated epitopes, cells were fixed with 1% paraformaldehyde at RT for 10 min. Then, cells were washed with a flow cytometry (FCM) buffer (PBS pH 7.2, 0.5% bovine serum albumin (BSA), 2 mM EDTA) and incubated with a mix of conjugated antibodies for 30 min at 4 °C in the dark. When a wide range of markers were analysed, we used a kit containing a combination of different conjugated monoclonal antibodies (CD90-FITC, CD105-PE, CD29-APC, CD44-FITC, CD166-PE, CD73-APC) (Immunostep, MCK-50T). Later, phycoerythrin (PE)-conjugated anti-human CD105 (Immunostep, 105PE-100T) and allophycocyanin (APC)-conjugated anti-human CD166 (Immunostep, 1399990314) were chosen for cell characterization. After the antibody incubation, the cells were washed twice and resuspended in 100–200 μl FCM buffer. For the detection of membrane and intracellular epitopes, the cells were collected, washed twice with PBS and fixed with 1% paraformaldehyde at RT for 10 min. Then, the cells were washed twice with PBS and permeabilized with 900 μl chilled (−20 °C) methanol (VWR Chemicals) for 30 min at 4 °C. Next, two washes with PBS were performed and the cells were kept in FCM buffer for 30 min prior to the antibody incubation. After being pelleted at 600 g for 7 min at 4 °C, the cells were resuspended in 100 μl of cold staining buffer with unconjugated monoclonal mouse anti-collagen II antibody (Invitrogen, Thermo Fisher Scientific, MA5-12789) or with APC-conjugated anti-human Cx43 antibody (R&D Systems, FAB7737A) for 30 min at 4 °C in the dark. When unconjugated primary antibodies were used, after two washes with the staining buffer, the cells were incubated with R-phycoerythrin (RPE)-conjugated goat anti-mouse antibody (DAKO, R0480) for 30 min at 4 °C in the dark. As a final step, the cells were washed twice with cold FCM buffer and resuspended in 100–200 µl of the same buffer until analysis.

#### Flow cytometry analysis

Cells were analysed on a BD Accuri C6 (Becton Dickinson,) and/or a BD FACSCalibur™ (Becton Dickinson) flow cytometer. Cell debris was discriminated by the forward scatter (FSC) and side scatter (SSC) properties of the cells, and cell aggregates were removed from the analysis by the selection of single cells using the area versus high signal of FSC, as described in Supplementary Fig. 6a. Between 5,000 and 20,000 events were collected in the single scatter gate region and data were analysed with FCS Express 6 Flow software (De Novo Software). The level of positive staining was expressed as the median fluorescence intensity or as a stain index by weighting the fluorescence with the standard deviation of each sample. Negative controls were unlabelled cells. Compensation was performed using antibody capture beads (the compensation matrix can be found in Supplementary Fig. 6b). Gates were placed based on single-labelled controls and by establishing 0.1% as the cutoff point.

#### Senescence-associated β-galactosidase activity

SAβG activity was assessed by flow cytometry with the fluorogenic β-galactosidase substrate di-β-D-galactopyranoside (FDG; Invitrogen, Thermo Fisher Scientific), which is hydrolyzed by cell endogenous β-galactosidase to fluorescein (FITC). Cells were collected and incubated in staining medium (PBS, 4% FBS, 10 mM HEPES, pH 7.2) at 37 °C for 10 min. The β-galactosidase assay was started by adding pre-warmed 2 mM FDG to the cell suspension and performing a 3-min incubation at 37 °C in the dark. The reaction was stopped with the addition of ice-cold staining medium to the cells, which were kept on ice and protected from light until analysis.

#### Immunofluorescence

Cells were fixed with 2% (w/v) paraformaldehyde (Sigma-Aldrich) in PBS for 10 min at RT and then incubated for an additional 10 min at RT with 0.1 M glycine (Sigma-Aldrich). Cell membranes were permeabilized by incubation with 0.2% Triton X-100 (Sigma-Aldrich) in PBS for 10 min at RT. After a PBS wash, cells were incubated for 30 min at RT with 1% BSA (Sigma-Aldrich) in PBS supplemented with 0.1% (v/v) Tween-20 (PBST; Sigma-Aldrich). Primary antibodies were diluted in 1% BSA in PBST and incubated for 1 h at RT. Three 10-min washes with PBS were performed and the cells were incubated with secondary antibody (diluted in 1% BSA in PBST) for 1 h at RT in the dark. The cells were washed three times with PBS, and nuclei were stained with 1 µg/ml 4',6-diamidino-2-phenylindole dihydrochloride (DAPI; Sigma-Aldrich) for 4 min at RT in the dark. After three PBS washes, the coverslips were mounted with a drop of glycergel aqueous mounting medium (Dako) on a glass microscope slide. The following primary antibodies were used: anti-Cx43 (Sigma-Aldrich, C6129), anti-collagen II (Invitrogen, Thermo Fischer Scientific, MA5-12789) and anti-vimentin (Santa Cruz Biotechnology, sc-373717). Goat anti-rabbit FITC-conjugated (F-2765) and goat anti-mouse Alexa 594-conjugated (A-11032) secondary antibodies were used (both from Invitrogen, Thermo Fisher Scientific). Fluorescence was analysed by using ImageJ software version 1.48 and is shown as the corrected total cell fluorescence (CTCF).

#### Immunohistochemistry

Acetone-fixed 4-µm-thick sections of cell micromasses were rinsed twice with PBS and non-specific endogenous peroxidase activity was quenched with a 3% (v/v) hydrogen peroxidase ($H_2O_2$) solution (Roche) for 10 min. The samples were washed twice with PBS and primary antibody was applied for 1 h at RT. The slides were then washed three times with PBS and incubated with OptiView HQ Universal Linker (Roche) for 10 min at RT. After three additional washes with PBS, the slides were incubated with OptiView HRP Multimer (Roche) for 8 min at RT. After two more washes with PBS, peroxidase activity was developed using a 0.02% $H_2O_2$ and 0.1% DAB solution. The sections were then washed in distilled water and, in some cases, counterstained with Gill's hematoxylin (Merck Millipore). As a final step, the slides were gradually dehydrated with alcohol, cleared with xylene (PanReac AppliChem) and mounted with DePeX (BDH Gun®, VWR).

#### Immunohistochemistry analysis

An in-house developed MATLAB program was used to quantify the immunohistochemistry images. This program first splits RGB images into single channels and applies a three-level automatic thresholding to the green channel using Otsu's method[116]. Then, it segments the images into four discrete classes using the threshold levels. Finally, the total pixels are obtained from the segmented images belonging to the three darker classes by discarding the lightest level in the image (background).

#### Quantitative PCR

Total RNA was isolated from cells using TRIzol™ reagent (Invitrogen, Thermo Fisher Scientific) according to the manufacturer's instructions. The RNA was treated with RNase-free DNase (Invitrogen, Thermo Fisher Scientific) to ensure the degradation of DNA in the samples. One microgram of total RNA per reaction was used to synthesize cDNA with the SuperScript® VILO™ cDNA Synthesis Kit as instructed by the manufacturer (Invitrogen, Thermo Fisher Scientific). Quantitative PCR was performed with the Applied Biosystems™ Power-UP™ SYBR™ Green Master Mix from Applied Biosystems on a real-time PCR instrument (LightCycler® 480 System, Roche) using the primers listed in Supplementary Table 1.

### Cell transfection

The T/C-28a2 chondrocyte cell line was transfected with a pIRESpuro2 plasmid construct (Clontech) containing the human Cx43 sequence, kindly provided by Arantxa Tabernero (Institute of Neuroscience of Castilla y León, University of Salamanca, Salamanca, Spain). Briefly, $10^6$ cells were electroporated with either 3 µg of pIRES-Cx43 plasmid or the pIRESpuro2 plasmid as control. Electroporation was performed with the Amaxa® Cell Line Nucleofector® Kit V (Lonza) in a Nucleofector™ 2b device (Lonza) following the manufacturer's instructions. Transfected cells were selected by supplementing the culture medium with 0.1 µg/ml puromycin dihydrochloride (Tocris).

### CRISPR/Cas9 system

T/C-28a2 cells were electroporated with the Amaxa® Cell Line Nucleofector® Kit V (Lonza) in a Nucleofector™ 2b device with 2 µg of a transitory all-in-one CRISPR expression vector (modified version of Addgene plasmid #48138) containing the VP12 high-fidelity Cas9 enzyme (derived from Addgene plasmid #72247) linked to a GFP marker via a 2A-self-cleaving peptide, with a guide targeting 20 nucleotides near the initial part of exon 2 of the Cx43 gene (5′-ACAGCGGTTGAGTCAGCCTG-3′). Single electroporated cells were seeded into a 96-well plate and $GFP^+$ clones were selected and expanded in conditioned medium (50% v/v medium from non-transfected T/C-28a2 cells). Dot blot and western blot analyses to check Cx43 levels were used to select the clones for further study.

### Statistical analysis

Data were analysed using GraphPad Prism software (version 5.00). Mean analysis of two groups was performed using either the Student's $t$-test or the Mann–Whitney $U$-test, whereas the mean of more than two groups was analysed with the one-way ANOVA with Tukey post hoc test. $P < 0.05$ was considered statistically significant.

#### Acknowledgements

This work was supported in part through funding from the Spanish Society for Rheumatology (SER; FER 2013) and Spanish Foundation for Research on Bone and Mineral Metabolism (FEIOMM), grant PRECIPITA-2015-000139 from the FECYT-Ministry of Economy and Competitiveness (to M.D.M.), grant PI16/00035 from the Health Institute 'Carlos III' (ISCIII, Spain), the European Regional Development Fund, 'A way of making Europe' from the European Union (to M.D.M.) and a grant from Xunta de Galicia IN607B 2017/21 (to M.D.M.) and pre-doctoral fellowship to M.V.-E. T.A. acknowledges support from Instituto de Salud Carlos III grants PI16/00772 and CPII16/00042, co-financed by the European Regional Development Fund (ERDF). We thank members of the CellCOM group for helpful technical suggestions, Ana González Sanchez for the Cx43 vector, María Dolores Álvarez Alvariño and Jesús Loureiro for generously collecting tissue samples in the operating room after surgery, Ángel Concha and the Biobank of A Coruña, María Vázquez and Beatriz Lema for tissue processing and micromass sectioning. Moisés Blanco for helpful advice for the statistical analysis of experimental data.

#### Author details

[1]CellCOM research Group, Instituto de Investigación Biomédica de A Coruña (INIBIC), Servizo Galego de Saúde (SERGAS), Universidade da Coruña (UDC), Xubias de Arriba, 84, 15006 A Coruña, Spain. [2]Service of Neurobiology Research, Ramón y Cajal University Hospital (IRYCIS), Madrid, Spain. [3]Centre for Medical Informatics and Radiological Diagnosis, Universidade da Coruña, A Coruña, Spain. [4]Bone and Joint Research Unit, Rheumatology Department, IIS-Fundación Jiménez Díaz UAM, Madrid, Spain. [5]Translational Molecular Pathology Research Group, Vall d'Hebron Research Institute (VHIR), Universitat Autònoma de Barcelona, CIBERONC, 08035 Barcelona, Spain. [6]Departamento de Bioquímica y Biología Molecular, Instituto de Neurociencias de Castilla y León (INCYL), Universidad de Salamanca, Salamanca, Spain. [7]Department of Pathology, School of Medicine, Wayne State University, Detroit, USA. [8]Department of Orthopaedic Surgery and Traumatology, Complexo Hospitalario Universitario de Santiago de Compostela (CHUS), Universidade de Santiago de Compostela (USC), Choupana s/n, 15706 Santiago de Compostela, Spain. [9]Flow Cytometry Core Technologies, UCD Conway Institute, University College Dublin, Dublin, Ireland

#### Author contributions

M.V.-E. designed and performed the experiments, analysed the data and prepared the figures. A.V.-V. and A.G.-C. performed experiments and assisted with cell culture and chondrocyte isolation from patients' cartilage. C.L.P. and R.L.-C. provided drugs, chondrocytes and hMSCs from healthy donors. V.M. analysed data. T.A., M.K. and A.T. provided vectors and technical guidance. E.F. and J.R.C. contributed clinical and technical advice. J.R.C. also provided cartilage tissue from patients and healthy donors. A.B. designed, performed and helped with various aspects of the flow cytometry experiments and analysis. M.D.M. conceived, directed and supervised the study. M.V.-E. and M.D.M. wrote the manuscript with input from all co-authors. All authors reviewed the manuscript.

#### Conflict of interest

The authors declare that they have no conflict of interest.